\documentclass[preprint, 3p]{elsarticle}

\usepackage{amsmath,amsxtra,amssymb,latexsym,amscd,amsthm}
\usepackage{enumerate}
\usepackage{graphics}

\usepackage[prependcaption,colorinlistoftodos]{todonotes}

\newtheorem{theorem}{Theorem}[section]
\newtheorem{lemma}[theorem]{Lemma}
\theoremstyle{definition}
\newtheorem{definition}[theorem]{Definition}
\newtheorem{example}[theorem]{Example}
\newtheorem{proposition}[theorem]{Proposition}
\newtheorem{remark}[theorem]{Remark}
\newtheorem{corollary}[theorem]{Corollary}

\journal{Systems $\&$ Control Letters}

\DeclareMathOperator{\col}{col}

\DeclareMathOperator{\conv}{conv}

\let\leq\leqslant
\let\geq\geqslant
\let\emptyset\varnothing

\newcommand{\calW}{\ensuremath{\mathcal{W}}}

\newcommand{\hatM}{\ensuremath{\hat{M}}}

\newcommand{\barx}{\ensuremath{\bar{x}}}

\newcommand{\barG}{\ensuremath{\bar{G}}}

\newcommand{\bmat}{\begin{matrix}}
\newcommand{\emat}{\end{matrix}}
\newcommand{\bbm}{\begin{bmatrix}}
\newcommand{\ebm}{\end{bmatrix}}
\newcommand{\bpm}{\begin{pmatrix}}
\newcommand{\epm}{\end{pmatrix}}
\newcommand{\bse}{\begin{subequations}}
\newcommand{\ese}{\end{subequations}}
\newcommand{\beq}{\begin{equation}}
\newcommand{\eeq}{\end{equation}}
\newcommand{\ben}{\begin{enumerate}}
\newcommand{\een}{\end{enumerate}}
\newcommand{\bit}{\begin{itemize}}
\newcommand{\eit}{\end{itemize}}
\newcommand{\bthe}{\begin{theorem}}
\newcommand{\ethe}{\end{theorem}}
\newcommand{\blem}{\begin{lemma}}
\newcommand{\elem}{\end{lemma}}
\newcommand{\bex}{\begin{example}}
\newcommand{\eex}{\end{example}}
\newcommand{\bas}{\begin{assumption}}
\newcommand{\eas}{\end{assumption}}
\newcommand{\bre}{\begin{remark}}
\newcommand{\ere}{\end{remark}}
\newcommand{\bcor}{\begin{corollary}}
\newcommand{\ecor}{\end{corollary}}

\newcommand{\pset}[1]{\ensuremath{\{#1\}}}
\newcommand{\set}[2]{\ensuremath{\{#1\mid #2\}}}

\newcommand{\norm}[1]{\ensuremath{\| #1 \|}}

\newcommand{\R}{\ensuremath{\mathbb R}}

\newcommand{\BP}{\noindent{\bf Proof. }}
\newcommand{\EP}{\hspace*{\fill} $\blacksquare$\bigskip\noindent}

\begin{document}

\begin{title}
{On the existence, uniqueness and nature of Carath\'eodory and Filippov solutions for bimodal piecewise affine dynamical systems}
\end{title}

\begin{frontmatter}

\author[kanat-thuan,thuan]{L. Q. Thuan}\ead{t.q.le@rug.nl}
\author[kanat-thuan,kanat]{M. K. Camlibel}\ead{m.k.camlibel@rug.nl}

\address[kanat-thuan]{Johann Bernoulli Institute for Mathematics and Computer Science, University of Groningen, P.O. Box 800, 9700 AV Groningen, The Netherlands}

\address[thuan]{Department of Mathematics, Quy Nhon University, 170 An Duong Vuong, Quy Nhon, Binh Dinh, Vietnam}

\address[kanat]{Department of Electronics and Communication Engineering, Dogus University, Kadikoy 34722, Istanbul, Turkey}

\begin{abstract} 

In this paper, we  deal with the well-posedness (in the sense of existence and uniqueness of solutions) and nature of solutions for discontinuous bimodal piecewise affine systems in a differential inclusion setting. First, we show that the conditions guaranteeing uniqueness of Filippov solutions in the context of general differential inclusions are quite restrictive when applied to bimodal piecewise affine systems. Later, we present a set of necessary and sufficient conditions for uniqueness of Filippov solutions for bimodal piecewise affine systems. We also study the so-called Zeno behavior (possibility of infinitely many switchings within a finite time interval) for Filippov solutions.
\end{abstract}

\begin{keyword}
Piecewise affine systems, well-posedness, existence and uniqueness of solutions, Carath\'eodory solutions, Filippov solutions, one-sided Lipschitz condition.
\end{keyword}
\end{frontmatter}

\section{Introduction}

A piecewise affine dynamical system is a special type of finite-dimensional, nonlinear input-state-output systems  with the distinguishing feature that the functions representing the system of differential equations and output equations are piecewise affine functions. Such  systems arise in various contexts of system and control theory such as variable structure systems \cite{utkin:92}, bang-bang control \cite{lasalle:59}, and linear relay systems \cite{pogromsky:03, lootsma:99}.

Piecewise affine functions which describe the dynamics of a piecewise affine dynamical system are not necessarily continuous. As such, piecewise affine dynamical systems form a subclass of discontinuous dynamical systems (see \cite{cortes:09} as an excellent survey). An immediate consequence of discontinuous dynamics is that the existing results of mainstream {\em smooth\/} nonlinear systems and control theory (see e.g. \cite{arjan-book}) cannot be indiscriminately applied to piecewise affine dynamical systems. This departure from {\em smooth\/} systems begins already from the definition of a notion of solution. Indeed, meaning of a solution of a differential equation given by continuous functions is rather straightforward whereas it becomes a much more complicated matter in the absence of continuity. 

The typical framework to deal with discontinuous dynamical systems is the framework of differential inclusions (see e.g. \cite{smirnov:02}). Roughly speaking, one replaces a differential equation with discontinuous right-hand side (see e.g. Filippov's seminal work \cite{filippov:88}) by a differential inclusion given by a set-valued mapping. There are several ways of defining a set-valued mapping (and hence a differential inclusion) for a differential equation with discontinuous right-hand side. Each of these leads to a different solution concept (see e.g. \cite{bacciotti:03, cortes:09} for details) such as Carath\'eodory, Krasovskii, Filippov, and Euler solutions. A good deal of the literature on differential inclusions is devoted to the investigation of existence and uniqueness of solutions. Typically, existence of solutions is guaranteed by less restrictive conditions than those for uniqueness. 

In this paper, we focus on a particular class of piecewise affine dynamical systems, namely bimodal piecewise affine systems without external inputs. The main goal of the paper is to investigate existence, uniqueness, and nature of solutions (in the sense Carath\'eodory and Filippov) for this class of systems. It turns out that existence of Filippov solutions immediately follows  from the existing results for general differential inclusions. However, existing conditions ensuring uniqueness for general differential inclusions are quite restrictive in the context of piecewise affine dynamical systems (see Theorem~\ref{thrm:main0} in Section~\ref{s:bimodal}). Motivated by this fact, we turn our attention to tailor-made conditions for bimodal systems. The main results of this paper are a set of necessary and a set of sufficient conditions for uniqueness of Filippov solutions of bimodal systems. Furthermore, these results provide necessary and sufficient conditions for the existence and uniqueness of Filippov solutions for bimodal piecewise linear systems.

A curious phenomenon in the context of discontinuous dynamical systems is the so-called Zeno behavior (see e.g. \cite{zhang:01, johansson:99}) which refers to infinitely many switchings in a finite time interval. Presence of such behavior causes serious difficulties not only in analysis and design but also in simulation of such systems. As a by-product of our main results, we obtain conditions under which Zeno behavior is ruled out for bimodal piecewise affine systems.

Well-posedness of piecewise affine dynamical systems has received considerable attention in the last two decades. In \cite{imura:00}, the authors consider bimodal piecewise linear systems. They work with what we call {\em forward\/} Carath\'eodory solutions and provide necessary and sufficient conditions for existence and uniqueness of these solutions. Forward Carath\'eodory solutions rule out the possibility of left accumulation points for switching instance by their very definition. In this paper, we consider more general bimodal systems, namely bimodal piecewise affine systems. Also we work not only with forward Carath\'eodory solutions but also with Filippov solutions. As such, the main result of \cite{imura:00} becomes a special case of our main results. In \cite{lootsma:99}, well-posedness of linear relay systems was addressed for forward Carath\'eodory solutions and sufficient conditions for uniqueness were presented. A linear relay system with a single relay boils down to a bimodal piecewise affine system as studied in this paper (see Example~\ref{ex:relay}). The results presented in this paper show that the very same conditions of \cite{lootsma:99} ensure uniqueness of Filippov solutions for this case. The paper \cite{pogromsky:03} studied linear relay systems with a single relay and provided sufficient conditions for the uniqueness of Filippov solutions. Also the results of \cite{pogromsky:03} can be recovered from our main results. Another related paper is \cite{camlibel:08} which considers Filippov solutions bimodal piecewise linear systems. The results of \cite{camlibel:08} can also be recovered as a special case from our main results  (see Corolloary~\ref{cor:1}). 

Zeno behavior of systems that are closely related to piecewise affine dynamical systems has been considered in \cite{shen:05} for a class of linear complementarity systems, in \cite{camlibel:06} for conewise linear systems, in \cite{schumacher:09} for linear relay systems with a single relay, and in \cite{thuan:11} for continuous bimodal piecewise affine systems. 

The paper is organized as follows. In Section~\ref{s:bimodal}, we introduce the object of the study in this paper, i.e. bimodal piecewise affine dynamical systems in a differential inclusion setting. We also present two examples of such systems and define what a solution means for such systems. This is followed by a discussion on existence of solutions as well as a discussion of the restrictiveness of conditions that guarantee uniqueness of Filippov solution of general differential inclusions when applied to bimodal systems. Section~\ref{s:unique} presents the main results related to the uniqueness of Filippov solutions whereas we investigate Zeno behavior of bimodal systems in Section~\ref{s:switching}. In Section~\ref{sec:proofs}, we present the proofs of the main results. Finally, the paper closes with the conclusions in Section~\ref{sec:conclusions}.

\section{Bimodal piecewise affine systems}\label{s:bimodal}

Consider the differential inclusion
\beq\label{eq:inc-1}
\dot x(t)  \in F (x(t))
\eeq
with
$$
F(x)=\begin{cases}  \{A_1x+e_1\} &\hskip-0.3cm \text{ if }\ c^T x +f< 0 \\ \{A_1x+e_1, A_2x+e_2\} &\hskip-0.3cm  \text{ if } \ c^T x+f = 0\\  \{A_2x+e_2\} &\hskip-0.3cm \text{ if } \ c^T x +f > 0\end{cases}
$$
where $x\in\R^n$, $A_1, A_2 \in \R^{n\times n}$, $e_1, e_2, c \in \R^n$ and $f\in \R$. Also consider the convexified differential inclusion
\beq\label{eq:inc-2}
\dot x(t)  \in G(x(t))
\eeq
with
$$
G (x)=\begin{cases} \{A_1x+e_1\} &\text{ if }\ c^T x +f< 0 \\ \conv(\{A_1x+e_1, A_2x+e_2\}) & \text{ if } \ c^T x+f = 0\\  \{A_2x+e_2\} &\text{ if } \ c^T x +f > 0\end{cases}
$$
where $\conv$ stands for the convex hull.

Throughout the paper, we call the systems of the form \eqref{eq:inc-1} and \eqref{eq:inc-2} {\em bimodal piecewise affine systems}. In the sequel, we investigate existence and uniqueness of different kinds of solutions of bimodal piecewise affine systems. 

Before elaborating on the solution concepts for these systems, we provide some examples of bimodal piecewise affine systems. 

The first class of examples consists of linear systems with ideal relay elements which serve as an idealized models of Coulomb friction, bang-bang control, etc.
\begin{example} \label{ex:relay}(Linear relay systems \cite{pogromsky:03,lootsma:99}) Consider the linear relay system
\begin{align*}
\dot x(t)&= Ax(t)+bu(t) \\ y(t)&= c^T x(t)\\
u(t) &\in -\text{sgn}(y(t))
\end{align*} 
where $\mathrm{sgn}$ is the set-valued relay function  defined by
$$
\mathrm{sgn}(y)=\begin{cases}
\{-1\}&\text{if }y<0\\
[-1,1]&\text{if }y=0\\
\{1\}&\text{if }y>0.
\end{cases}
$$
Clearly, such a linear relay system is a bimodal piecewise affine system of the following form
\begin{equation*}
\dot x(t) \in\begin{cases}  \{Ax(t)+b\} & \text{ if }\ c^T x(t)< 0 \\ \conv(\{Ax(t)+b, Ax(t)-b\}) &  \text{ if } \ c^T x(t)= 0\\  \{Ax(t)-b\} & \text{ if } \ c^T x(t) > 0.\end{cases}
\end{equation*}
\end{example}

Note that linear relay systems are particular cases of \eqref{eq:inc-2} where $A_1=A_2$ and $f=0$. Well-posedness of such systems were studied in \cite{pogromsky:03} and  \cite{lootsma:99}. Another example of bimodal piecewise affine systems is the following water tank system.

\begin{example}\label{ex:2} Consider the two tank system depicted in Figure~\ref{fig-pca}. 
\begin{figure}[!ht]
\centering
\includegraphics[height=6cm,width=8cm]{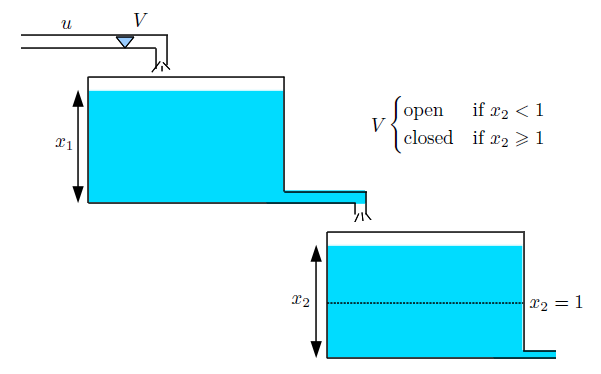}
\caption{Water level regulator}\label{fig-pca}
\end{figure}

The deviations of the water level from the bottom of the first and second tank are denoted by $x_1$ and $x_2$, respectively. Let $u$ be the constant flow of water into the first tank. The valve $V$ is opened if $ x_2 < 1$ and closed if $x_2\geqslant 1$. By defining the state $x =\col(x_1, x_2)$ and taking all involved parameters unity, one obtains the following equations describing the dynamics of the system
\begin{align*}\label{eq:sys:04}
\dot x& = \begin{bmatrix} -1 &0\\ 1&-1\end{bmatrix} x + \begin{bmatrix} u \\ 0 \end{bmatrix} \quad\text{ if }  x_2 -1 < 0 \\[5mm] \dot x& = \begin{bmatrix} -1 &0\\ 1&-1\end{bmatrix} x + \begin{bmatrix} 0 \\ 0 \end{bmatrix} \quad \text{ if }  x_2 -1 \geqslant 0 
\end{align*}
As such, this system is of the form of a bimodal piecewise affine system \eqref{eq:inc-1}.
\end{example}

We now turn our attention to formalizing what will be meant by a solution of the system \eqref{eq:inc-1}. There are many ways of defining a solution for a differential inclusion (see e.g. \cite{bacciotti:03,cortes:09}).
In this paper, we focus on Carath\' eodory and Filippov solution concepts.

\begin{definition} \label{def:solution} An absolutely continuous function $x: \R \to \R^n $ is said to be a solution of the system \eqref{eq:inc-1} for the initial state $\xi$ in the sense of 
\begin{itemize}
\item {\em Carath\'{e}odory} if $x(0) = \xi $ and $x$ satisfies the differential inclusion \eqref{eq:inc-1} for almost all $ t \in \R$.
\item {\itshape  forward Carath\'{e}odory}  if it is a solution in the sense of Carath\'{e}odory, and for each $t^* $ there exist $i\in\{1,2\}$ and $\epsilon_{t^*}> 0 $ such that 
\begin{equation}
\dot x(t) = A_ix(t) + e_i\text{ and } (-1)^{i-1}[ c^Tx(t)+f]  \leqslant 0 \label{eq:fc}
\end{equation}
for all $t\in ( t^*, t^*+\epsilon_{t^*})$. 

\item {\itshape  backward Carath\'{e}odory} if it is a solution in the sense of Carath\'{e}odory, and for each $t^* $ there exist $i\in\{1,2\}$ and $\epsilon_{t^*}> 0 $ such that
\begin{equation}
 \dot x(t) = A_ix(t) + e_i\text{ and }  (-1)^{i-1}[ c^Tx(t) +f]  \leqslant 0 \label{eq:bc}
\end{equation}
for all $t\in (t^*-\epsilon_{t^*}, t^*)$.
\item {\itshape Filippov} if $ x(0) = \xi $ and $x$ satisfies the convexified differential inclusion \eqref{eq:inc-2} for almost all $ t \in \R$.
\end{itemize}
\end{definition}
Clearly, every Carath\'{e}odory solution is a Filippov solution since $F(x)\subseteq G(x)$ for all $x\in\R^n$. However, not every Filippov solution is a Carath\'{e}odory solution in general.

When the right hand side of \eqref{eq:inc-1} is single-valued and hence is Lipschitz continuous, that is 
the implication 
\begin{align}
c^T x +f = 0\quad \implies \quad A_1x +e_1 = A_2x+e_2 \label{cc}
\end{align}
holds, Carath\'eodory and Filippov solutions coincide. In this case, existence and uniqueness of solutions are guaranteed by the theory of ordinary differential equations. 

In general, existence of solutions of the differential inclusion \eqref{eq:inc-1} readily follows from the theory of differential inclusions (see e.g. \cite[Theorem 2.7.1]{filippov:88}). 

\begin{proposition}
There exists a solution of the differential inclusion \eqref{eq:inc-1} in the sense of Filippov for each initial state. 
\end{proposition}

In the sequel we focus on the uniqueness of Filippov solutions for the system \eqref{eq:inc-1}. 

\begin{definition} We say that a Filippov solution for the initial state $\xi$ is {\em right-unique\/} ({\em left-unique\/}) if for any Filippov solution $x'$ for the initial state $\xi$ there exists $\epsilon >0$ such that $x(t)=x'(t)$ for all $t\in [0,\epsilon)$ ($t\in (-\epsilon, 0]$).
\end{definition}

The main goal of the paper is to present necessary and/or sufficient conditions for uniqueness of Filippov solutions that are tailored to bimodal piecewise affine systems of the form \eqref{eq:inc-1}. To motivate these new conditions, we first review one of the most typical uniqueness conditions that is employed in the literature of (general) differential inclusions and discuss its limitations for bimodal piecewise affine systems.
 
A set-valued mapping $H:\R^n\rightrightarrows\R^n$ is said to be {\em one-sided Lipschitz\/} (see e.g. \cite{cortes:09}) if there exists a number $L$ such that 
\beq\label{eq:lipc-1}
(x_1-x_2)^T(y_1-y_2) \leqslant L ||x_1-x_2||^2
\eeq
for all $x_1$, $x_2$ belonging to the domain of $H$, $y_1\in H(x_1)$ and $y_2\in H(x_2)$. 

The following theorem presents necessary and sufficient conditions for one-sided Lipschitzness of the set-valued mapping $G$.

\begin{theorem}\label{thrm:main0} The set-valued mapping $G$ is one-sided Lipschitz if and only if there exist a vector $g\in \R^n$ and a number $\mu \geqslant 0$ such that 
$$
A_1 -A_2 = g c^T  \text{ and  } e_1 - e_2 = f g+\mu c.
$$
\end{theorem}

\BP For the `if' part, suppose that we have $A_1 -A_2 = g c^T, e_1 - e_2 = fg+\mu c$ where $\mu\geq 0$. Then, the piecewise affine function which is defined by
$$\tilde G(x) =  \begin{cases} A_1x + e_1-\dfrac{\mu c}{2}& \text{if} \ c^T x+f \leqslant 0\\[2mm]  A_2x +e_2+\dfrac{\mu c}{2}& \text{if} \ c^T x+f \geqslant  0\end{cases}
$$
is continuous, so that it is globally Lipschitz continuous (see \cite{facchinei:02}).  Observe that
$
G(x) = \tilde G(x) + [1-2\lambda(x)]\dfrac{\mu c}{2} 
$
where 
$$
\lambda (x) = \begin{cases}\{0\} &\text{if } \ c^T x +f <0 \\ [0,1] &\text{if } \ c^T x +f =0\\ \{1\} &\text{if } \ c^T x +f >0. \end{cases}
$$
Thus, for any $x_i\in \R^n$ and  $y_i\in G(x_i)$, there exists $\bar y_i \in \tilde G(x_i)$ such that 
$ y_i=\tilde{y}_i+(1-2\lambda_i)\frac{\mu}{2}c$ where $\lambda_i\in\lambda(x_i)$, and then 
$$
(x_1-x_2)^T(y_1 -y_2) =  (x_1-x_2)^T\big [(\tilde y_1 - \tilde y_2) - \mu(\lambda_1-\lambda_2)c\big ].
$$
This together with the observation that $\mu\geq 0$ and $(\lambda_1-\lambda_2)(x_1 -x_2)^T c \geqslant 0$ implies that
$$
(x_1-x_2)^T(y_1 -y_2)\leq (x_1-x_2)^T(\tilde y_1 - \tilde y_2).
$$
On the other hand, the  Cauchy-Schwarz inequality and Lipschitzness of $\tilde G$ implies $(x_1-x_2)^T(\tilde y_1 - \tilde y_2) \leqslant L \norm{x_1-x_2}^2$ where $L$ denotes a Lipschitz constant of $\tilde G$. Thus, we have 
$$(x_1-x_2)^T(y_1 - y_2) \leqslant L \norm{x_1-x_2}^2$$
for all $x_1, x_2 \in \R^n$ and $y_i \in G(x_i)$, and hence $G$ is one-sided Lipschitz.

For the `only if' part, we suppose that $G$ is one-sided Lipschitz. Let 
$
\Sigma_{-} =\{x \ |\ c^Tx+f \leqslant 0\}$ and $ \Sigma_+=\{x \ | \ c^Tx +f \geqslant 0\}.
$
Let $x_1 \in \Sigma_{-}, x_2\in \Sigma_+ $, and let $\bar x$  be such that $c^T\bar x + f =0$. For $\alpha \in (0, 1]$, define 
$ x_1' = \alpha x_1 +(1-\alpha)\bar x $ and $ \  x_2' = \alpha x_2 +(1-\alpha)\bar x.$ Clearly, $x'_1 \in \Sigma_{-}$ and $ x'_2\in \Sigma_+$. Since $G$ is one-sided Lipschitz, one has
$$ 
[(A_1x_1'+e_1)-(A_2x_2'+e_2)]^T(x_1'-x_2') \leqslant L ||x_1'-x_2'||^2,
$$
or equivalently
$$
\big[\dfrac{(1-\alpha)}{\alpha} \big\{(A_1-A_2)\bar x +(e_1-e_2)\big\} + \big\{(A_1x_1+ e_1) -(A_2x_2+e_2)\big\}\big]^T(x_1-x_2) \leqslant L ||x_1-x_2||^2. 
$$
By taking sufficiently small $\alpha$, we obtain that 
$[(A_1-A_2)\bar x +(e_1-e_2)]^T(x_1-x_2) \leqslant 0
$
for all $x_1 \in \Sigma_{-}$ and $ x_2 \in \Sigma_+$. This implies that 
\begin{align}
(A_1-A_2)\bar x +(e_1-e_2)\in (\Sigma_{-}-\Sigma_+)^o \label{h}
\end{align}
for any $\bar x $ with $c^T\bar x +f =0$. Here the notation $S^o$ denotes the polar cone of the set $S$ that is $S^o=\set{y}{x^Ty\leq 0,\,\,\,\forall x\in S}$. Then, we get 
\begin{align*} (A_1-A_2)(\ker c^T) + (A_1-A_2)\bar x + (e_1-e_2)  \subseteq (\Sigma_{-}-\Sigma_+)^o =\{ \alpha c \ | \ \alpha \geqslant 0 \}\end{align*}
for fixed $\bar x$ satisfying $c^T\bar x +f =0$. Since the left hand side is an affine set and the right hand side is a cone, we can conclude that
$(A_1-A_2)(\ker c^T) =\{0\}.$
Hence, $A_1-A_2 = gc^T$ for some $g$. Then it follows from \eqref{h} that 
$e_1-e_2-fg \in (\Sigma_{-}- \Sigma_+)^o. $ 
Note that 
$
\Sigma_{-}- \Sigma_+ = \{ x \ | \ c^Tx \leqslant 0\}.
$
Hence, we have 
$
(\Sigma_{-}- \Sigma_+)^o = \{ \alpha c \ |\  \alpha \geqslant 0\}.
$
This means that $e_1-e_2 = fg +\mu c $ for some $\mu \geqslant 0$.
\EP

Theorem~\ref{thrm:main0} shows that we can employ one-sided Lipschitzness in order to conclude uniqueness of Filippov solutions of bimodal systems only under quite restrictive conditions. Note that these conditions are met for a linear relay system as in Example~\ref{ex:relay} only if $b=\alpha c$ for a nonnegative real number $\alpha$. Note also that the conditions of Theorem~\ref{thrm:main0} are never met for a two-tank system as in Example~\ref{ex:2}. Motivated by the restrictiveness of one-sided Lipschitzness, we investigate tailor-made uniqueness conditions for bimodal systems in the sequel.

\section{Uniqueness of solutions}\label{s:unique}

In this section, a set of necessary and a set of sufficient conditions for right-uniqueness of solutions of the bimodal piecewise affine systems \eqref{eq:inc-1} will be provided. These conditions are less restrictive than the conditions guaranteeing the one-sided Lipschitz condition. To do so, we need to introduce some nomenclature as follows. For a vector $v$, we write $v\succ 0$ if it is nonzero and the first nonzero entry is positive. We write $v \succeq 0$ if either $v=0$ or $v \succ 0$. Similarly, we write $v \prec 0$ when $-v \succ 0$ and $v \preceq 0$ when $-v \succeq 0$. The observability index of the pair $(c^T,A_i)$ is denoted by $h_i$, that is the largest integer such that the matrix $\text{col}(c^T, c^TA_i, \ldots, c^TA_i^{h_i})$ has full row rank. Note that for each $k\geq 1$ there exists a unique matrix $P_i^k\in\R^{k\times h_i}$ such that
\beq \label{eq:care}
\bbm
c^TA_i^{h_i+1}\\c^TA_i^{h_i+2}\\\vdots\\c^TA_i^{h_i+k}\ebm=P_i^k 
\bbm
c^T\\c^TA_i\\\vdots\\c^TA_i^{h_i}
\ebm.
\eeq

We now present the main results concerning the well-posedness of solutions of the bimodal piecewise affine system \eqref{eq:inc-1} in the three theorems below.

\begin{theorem} \label{t:main1} Let $h:=\min\{h_1,h_2\}$. Consider the statements:
\begin{enumerate}
\item  \label{t:main1:s1} Every Filippov solution of \eqref{eq:inc-1} is right-unique.

\item \label{t:main1:s3} Every Filippov solution of \eqref{eq:inc-1} is both a forward and backward Carath\'eodory solution.

\item \label{t:main1:s4} There exist an integer $k$ with $1\leqslant k \leqslant h+1$ and a $(k+1)\times (k+1)$ lower triangular matrix $M$ with positive diagonal elements such that 
$$
\hskip-13pt \begin{bmatrix}c^T\\ c^TA_1 \\ \vdots \\  c^TA_1^k\end{bmatrix} = M \begin{bmatrix}c^T\\ c^TA_2 \\ \vdots \\  c^TA_2^k\end{bmatrix}, \begin{bmatrix} f\\ c^Te_1 \\ \vdots \\ c^TA_1^{k-1}e_1\end{bmatrix} \succ M \begin{bmatrix} f\\ c^Te_2 \\ \vdots \\ c^TA_2^{k-1}e_2\end{bmatrix}.
$$ 
\item \label{t:main1:s4.5} There exists a $(h+1)\times (h+1)$ lower triangular matrix $M$ with positive diagonal elements such that
$$
\hskip-13pt \begin{bmatrix}c^T\\ c^TA_1 \\ \vdots \\  c^TA_1^h\end{bmatrix} = M \begin{bmatrix}c^T\\ c^TA_2 \\ \vdots \\  c^TA_2^h\end{bmatrix}, \begin{bmatrix} f\\ c^Te_1 \\ \vdots \\ c^TA_1^{h-1}e_1\end{bmatrix} = M \begin{bmatrix} f\\ c^Te_2 \\ \vdots \\ c^TA_2^{h-1}e_2\end{bmatrix}
$$
and either
$
h_1<h_2 \text{ and } c^TA_1^{h_1}e_1 -p_1^T{\bf e}_1^{h_1} >0
$
or
$
h_1>h_2 \text{ and } c^TA_2^{h_2}e_2 -p_2^T{\bf e}_2^{h_2} <0 $
where $p_i$ is uniquely determined from 
$$
c^TA_i^{h_i+1} = p_i^T T_i^{h_i}, i=1,2.
$$

\item \label{t:main1:s5} The observability indices $h_1$ and $h_2$ are the same and there exists a $(h+2)\times (h+2)$ lower triangular matrix $M$ with positive diagonal elements such that 
$$
\hskip-13pt \begin{bmatrix}c^T\\ c^TA_1 \\ \vdots \\  c^TA_1^{h+1}\end{bmatrix} = M \begin{bmatrix}c^T\\ c^TA_2 \\ \vdots \\  c^TA_2^{h+1}\end{bmatrix}, \begin{bmatrix} f\\ c^Te_1 \\ \vdots \\ c^TA_1^{h}e_1\end{bmatrix} = M \begin{bmatrix} f\\ c^Te_2 \\ \vdots \\ c^TA_2^{h}e_2\end{bmatrix}.
$$ 

\item \label{t:main1:s6} The following implication holds
$$
\begin{bmatrix}c^T\\ c^TA_1 \\ \vdots \\  c^TA_1^{h+1}\end{bmatrix}\xi +\begin{bmatrix} f\\ c^Te_1 \\ \vdots \\ c^TA_1^{h}e_1\end{bmatrix} = 0 \Rightarrow A_1\xi+e_1 = A_2\xi +e_2.
$$
\end{enumerate}
Then, the following implications hold:
\begin{enumerate}
\renewcommand{\theenumi}{\Alph{enumi}}
\renewcommand{\labelenumi}{\Alph{enumi}.}
\item\label{t:main1-a} $\ref{t:main1:s1} \Rightarrow (\ref{t:main1:s4} \text{ or }  \ref{t:main1:s4.5} \text{ or }\ref{t:main1:s5})$

\item\label{t:main1-b} $(\ref{t:main1:s1} \text{ and } \ref{t:main1:s5}) \Rightarrow \ref{t:main1:s6}$

\item\label{t:main1-c} $(\ref{t:main1:s5} \text{ and }  \ref{t:main1:s6}) \Rightarrow \ref{t:main1:s3} $

\item\label{t:main1-d} $(\ref{t:main1:s5} \text{ and } \ref{t:main1:s6}) \Rightarrow \ref{t:main1:s1}$
\end{enumerate}
\end{theorem}

A proof of this theorem will be presented in Section~\ref{sec:proofs}. Note that this theorem provides only a set of necessary and a set of sufficient conditions, but not necessary and sufficient conditions in general. The following example (see \cite[Eq's. (13) and (14)]{pogromsky:03}) illustrates the gap between the necessary and the sufficient conditions.

\begin{example} Consider the bimodal piecewise affine system \eqref{eq:inc-1} with $c^T=\begin{bmatrix} 1&0&0\end{bmatrix}$, $f=0$,
$$
A_1 = A_2 =\begin{bmatrix}0&1&0\\ 0&0&1 \\ 0&0&0 \end{bmatrix},\text{ and } e_1 = -e_2 =\begin{bmatrix}0\\0\\1 \end{bmatrix}.
$$
For this system, it can be verified that the third statement of Theorem \ref{t:main1} is satisfied with $k=3$. However, as it has been shown in \cite{pogromsky:03}, there are infinitely many Filippov solutions for the zero initial state.
\end{example}

However, the third statement of Theorem \ref{t:main1} with $k=1$ is sufficient for right-uniqueness of Filippov solutions as stated in the following.

\begin{theorem}\label{thrm:main2} If there exists a $2\times 2 $ lower triangular matrix $M$ with positive diagonal elements such that 
\begin{equation}\label{eq:thrm:main2}
\begin{bmatrix}c^T\\ c^TA_1\end{bmatrix} = M \begin{bmatrix}c^T\\ c^TA_2 \end{bmatrix}, \begin{bmatrix} f\\ c^Te_1\end{bmatrix} \succ M \begin{bmatrix} f\\ c^Te_2 \end{bmatrix}
\end{equation}
then right-uniqueness of Filippov solutions holds at any state of the state space $\R^n$.
\end{theorem}
\BP Due to \eqref{eq:thrm:main2}, for each state $\xi$ satisfying $c^T\xi + f =0$ at least one of the inequalities $c^TA_1 \xi + c^Te_1 >0$ or $c^TA_2\xi + c^Te_2 <0$ is satisfied. By \cite[Theorem 2.10.2]{filippov:88}, every Filippov solution is right-unique. 
\EP

Further, the third statement of Theorem \ref{t:main1} with $k=2$ is sufficient for right-uniqueness of Filippov solutions for some initial states.

\begin{theorem}\label{thrm:main3}  If there exists $3\times 3$ lower triangular matrix $M$ with positive diagonal elements such that 
\begin{equation}\label{eq:thrm:main3-0}
\begin{bmatrix}c^T\\ c^TA_1 \\  c^TA_1^{2}\end{bmatrix} = M \begin{bmatrix}c^T\\ c^TA_2 \\   c^TA_2^{2}\end{bmatrix}, \begin{bmatrix} f\\ c^Te_1 \\ c^TA_1e_1\end{bmatrix} \succ M \begin{bmatrix} f\\ c^Te_2 \\  c^TA_2 e_2\end{bmatrix}
\end{equation}
then  for the system \eqref{eq:inc-1}, right-uniqueness of Filippov solutions holds at all states from $\R^n\backslash \Omega$ where $\Omega$ is the set of all
 $\xi \in \R^n$ such that
$$
\begin{bmatrix} c^T\\ c^TA_j \\ c^TA_j^{2} \end{bmatrix}\xi + \begin{bmatrix} f\\ c^Te_j \\ c^TA_je_j \end{bmatrix}=0 \ \text{and} \ (-1)^{k}(c^TA_{k}^2\xi + c^T A_k e_k) <0 
$$
for some $k,j\in \{1,2\}, k\ne j$.  
\end{theorem}

A proof of this theorem will be given in Section~\ref{sec:proofs}.\\

Theorem~\ref{thrm:main2} and Theorem~\ref{thrm:main3} present two particular cases under which the third statement of Theorem~\ref{t:main1} becomes sufficient as well as necessary for right-uniqueness of Filippov solutions for bimodal systems. Another interesting particular case occurs when there are no affine terms in the dynamics, that is when $e_1=e_2=0$ and $f=0$. In this case, one can state necessary and sufficient conditions (see also \cite{camlibel:08}) as in the following.

\begin{corollary}\label{cor:1} Consider the system \eqref{eq:inc-1} with $e_1=e_2=0$ and $f=0$. Then, every Filippov solution of \eqref{eq:inc-1} is unique if and only if the following statements hold:
\begin{enumerate}
\item $h_1 = h_2$.
\item There exists an $(h_1+1) \times (h_1+1)$ lower triangular matrix $M$  with positive diagonal elements such that 
$$
\begin{bmatrix}c^T\\ c^TA_1 \\ \vdots \\  c^TA_1^{h_1}\end{bmatrix} = M \begin{bmatrix}c^T\\ c^TA_2 \\ \vdots \\  c^TA_2^{h_1}\end{bmatrix}.
$$ 
\item The following implication holds
$$
\begin{bmatrix}c^T\\ c^TA_1 \\ \vdots \\  c^TA_1^{h_1}\end{bmatrix}x =0 \Rightarrow A_1 x = A_2 x.
$$
\end{enumerate}
\end{corollary}

\BP
Note that the third and fourth statements of Theorem~\ref{t:main1} never holds as $e_1=e_2=0$ and $f=0$. Then, the first statement of Theorem~\ref{t:main1} holds if and only if the fifth and the sixth hold.\EP

\newcommand{\ts}{\ensuremath{t^*}}
\newcommand{\eps}{\epsilon}

\section{Switching behavior}\label{s:switching}
In this section, we investigate mode switching behavior of bimodal systems. We say that a time instant $\ts\in\R$ is a {\em non-switching time\/} for a Filippov solution $x$ if there exist an interval $(\ts-\eps,\ts+\eps)$ and an index $i$ with $i\in\{1,2\}$ such that
$$
\dot x(t) = A_ix(t) + e_i\text{ and } (-1)^{i-1}[ c^Tx(t)+f]  \leqslant 0 \label{eq:fc}
$$
for all $t\in(\ts-\eps,\ts+\eps)$. We say that a time instant $\ts\in\R$ is a {\em switching time\/} for a Filippov solution $x$ if $\ts$ is not a non-switching time for the same solution.

The distribution of the switching times along a solution is an important issue for various reasons. For instance, the so-called event-driven simulation methods (see e.g. \cite{Schaft:96}) may fail if the switching times accumulate around a point. This type of phenomenon is known as Zeno behavior in hybrid systems literature. 
We say that a time instant $\ts\in\R$ is a {\em left/right Zeno time\/} for a Filippov solution $x$ if for each $\eps>0$ the interval $(\ts,\ts+\eps)$/$(\ts-\eps,\ts)$ contains a switching time for the same solution.

A Fillipov solution will be called {\em left (right) Zeno-free\/} if there exists no left (right) Zeno time for it. Also, we say that the system \eqref{eq:inc-1} is {\em left (right) Zeno-free\/} if all its Fillipov solutions are left (right) Zeno-free. Further, we say that 
the system \eqref{eq:inc-1} is {\em Zeno-free\/} if it is both is left and right Zeno-free.

With these preparations, we can state the following sufficient condition for Zeno-freeness.

\bthe
Suppose that the observability indices $h_1$ and $h_2$ are the same. Let $h=h_1=h_2$. If there exists a $(h+2)\times (h+2)$ lower triangular matrix $M$ with positive diagonal elements such that 
$$
\hskip-13pt \begin{bmatrix}c^T\\ c^TA_1 \\ \vdots \\  c^TA_1^{h+1}\end{bmatrix} = M \begin{bmatrix}c^T\\ c^TA_2 \\ \vdots \\  c^TA_2^{h+1}\end{bmatrix}, \begin{bmatrix} f\\ c^Te_1 \\ \vdots \\ c^TA_1^{h}e_1\end{bmatrix} = M \begin{bmatrix} f\\ c^Te_2 \\ \vdots \\ c^TA_2^{h}e_2\end{bmatrix}
$$ 
and the implication
$$
\begin{bmatrix}c^T\\ c^TA_1 \\ \vdots \\  c^TA_1^{h+1}\end{bmatrix}\xi +\begin{bmatrix} f\\ c^Te_1 \\ \vdots \\ c^TA_1^{h}e_1\end{bmatrix} = 0 \Rightarrow A_1\xi+e_1 = A_2\xi +e_2
$$
holds, then the system \eqref{eq:inc-1} is Zeno-free.
\ethe

\BP
Note that the system \eqref{eq:inc-1} is Zeno-free if and only if any Filippov solution is both forward and backward Carath\'{e}odory solution. As such, the claim follows from Theorem~\ref{t:main1}.\ref{t:main1-c}.
\EP

\section{Proofs}\label{sec:proofs}

\subsection{Proof of Theorem~\ref{t:main1}}
To prove Theorem ~\ref{t:main1}, we first consider affine systems and introduce some notations. An affine system $\Sigma(A,e,c^T,f)$ is given by 
\begin{subequations}\label{eq:afsys}
\begin{align}
\dot x &= A x + e\\
y &= c^T x + f 
\end{align}
\end{subequations}
where $x\in\R^n$ is the state, $y\in\R$ is the output, and all involved matrices are of appropriate sizes. By $ x(t;\xi) $ and $ y(t;\xi)$, we denote the state and the output of the system \eqref{eq:afsys} for the initial state $\xi$, respectively. We define the sets
\begin{align*}
\mathcal W_\Sigma^{-} &:= \{\xi \ |\ \exists \epsilon  > 0 \ \text{ such that} \ y(t;\xi) < 0, \ \forall t \in (0, \epsilon)\},\\ 
\mathcal W_\Sigma^{0} &:= \{\xi \ |\ \exists \epsilon  > 0 \ \text{ such that} \ y(t;\xi) = 0, \ \forall t \in (0, \epsilon)\},\\
\mathcal W_\Sigma^{+} &:= \{\xi \ |\ \exists \epsilon  > 0 \ \text{ such that} \ y(t;\xi) > 0, \ \forall t \in (0, \epsilon)\}.
\end{align*}
Since $y(t;\xi)$ is a real-analytic function for each initial state $\xi$, its sign on a small time interval $(0,\epsilon)$ is completely determined by the values of its derivatives at $t=0$, i.e. $y^{(k)}(0;\xi)$ for $k=0,1,\ldots$. Note that 
$y^{(0)}(0;\xi)= c^T\xi +f$ and $y^{(k)}(0;\xi)= c^TA^k\xi + c^TA^{k-1}e$ for $ k \geqslant 1$. Let $\nu $ denote the observability index of the pair $(c^T,A)$. It can be verified that $c^TA^{\ell} \xi + c^TA^{\ell-1}e=0$ for all $\ell$ with $\nu+1\geqslant \ell \geqslant 1$ implies $ c^TA^{\ell} \xi + c^TA^{\ell-1}e=0$ for all $\ell\geq 1$. Together with the analyticity of the output $y(t;\xi)$, this observation  leads to the following immediate characterizations of the $\mathcal W$-sets:
\beq\label{e:prop:wch1}
\begin{pmatrix} \mathcal W_\Sigma^{-}\\ \mathcal W_\Sigma^{0} \\ \mathcal W_\Sigma^{+} \end{pmatrix} = \Big \{ \xi \Big | \  \begin{bmatrix} c^T\\ c^TA\\ \vdots \\ c^TA^{\nu+1} \end{bmatrix}\xi + \begin{bmatrix} f\\ c^Te \\ \vdots \\ c^TA^{\nu}e \end{bmatrix} \begin{pmatrix}\prec \\ = \\ \succ  \end{pmatrix} 0 \Big\}. 
\eeq
Note also that
\beq\label{e:prop:wch2}
\mathcal W_\Sigma^{- }  \cup \mathcal W_\Sigma^{0}  \cup \mathcal W_\Sigma^+  = \R^n. 
\eeq
In the sequel, we often use the $\calW$-sets corresponding to the two modes of the system \eqref{eq:inc-1}. For brevity, we denote $\Sigma_i = \Sigma_i(A_i, e_i,c^T, f)$ and define
\bse\label{e:w-sets-all}
\begin{gather}
\mathcal W_1^0:= \mathcal W_{\Sigma_1}^0  \quad \mathcal W_1^-:= \mathcal W_{\Sigma_1}^-   \quad \mathcal W_1:= \mathcal W_{\Sigma_1}^0 \cup \mathcal W_{\Sigma_1}^-  \\ 
 \mathcal W_2^0 := \mathcal W_{\Sigma_2}^0  \quad \mathcal W_2^+ := \mathcal W_{\Sigma_2}^+\quad \mathcal W_2 := \mathcal W_{\Sigma_2}^0 \cup  \mathcal W_{\Sigma_2}^+.
\end{gather}
\ese
We also define
$$
T_i^k := \begin{bmatrix} c^T\\ c^TA_i\\ \vdots \\ c^TA_i^k \end{bmatrix} \text{ and } \ {\bf e}_i^{k}:= \begin{bmatrix} f\\ c^Te_i \\ \vdots \\ c^TA_i^{k-1}e_i \end{bmatrix}
$$
for $k\geq 1$. With these preparations, we are ready to prove Theorem~\ref{t:main1}.

\subsubsection{Proof of Theorem~\ref{t:main1}.\ref{t:main1-a}}
Right-uniqueness of every Filippov solution of the system \eqref{eq:inc-1} implies that
\beq
\mathcal W_1^- \cap \mathcal W_2 = \emptyset\quad\text{and}\quad  \mathcal W_1 \cap \mathcal W_2^+ = \emptyset.
\eeq
In view of \eqref{e:prop:wch1} and \eqref{e:w-sets-all}, therefore, we have
\bse\label{e:T-h+1}
\begin{gather}
T_1^{h_1+1}\xi+{\bf e}_1^{h_1+1}\prec 0\implies T_2^{h_2+1}\xi+{\bf e}_2^{h_2+1}\prec 0\label{e:T-h+1.1}\\
T_2^{h_2+1}\xi+{\bf e}_2^{h_2+1}\succ 0\implies T_1^{h_1+1}\xi+{\bf e}_1^{h_1+1}\succ 0\label{e:T-h+1.2}
\end{gather}
\ese
as necessary conditions for right-uniqueness of every Filippov solution. Note that a consequence of the first implication is that
\beq\label{e:T-h}
T_1^{h}\xi+{\bf e}_1^{h}\prec 0\quad\implies\quad T_2^{h}\xi+{\bf e}_2^{h}\preceq 0.
\eeq
In order to formulate this condition in terms of the parameters of the system \eqref{eq:inc-1}, we invoke the following lemma which was proven in \cite{thuan:10}.

\begin{lemma}\label{le:04}
Let $P_i$ be an $m\times n$ matrix of full row rank and $q_i$ be an $m$-vector for $i=1,2$. Then, the following statements are equivalent:
\begin{enumerate}
\item $P_1x \prec q_1$ implies $P_2x \preceq q_2$.
\item $P_1x\prec q_1 $ implies $ P_2x\prec q_2 $.
\item  Either 
$$
P_1 = MP_2 \quad\text{and}\quad q_1 = Mq_2
$$
for some $m\times m$ lower triangular matrix $M$ with positive diagonal elements, or there exist $\ell$ with $ 1\leqslant\ell \leqslant m $ and $\ell\times \ell$ lower triangular matrix $M$ with positive diagonal elements such that  
$$
P_1^{[\ell]} = M P_2^{[\ell]}\quad\text{and}\quad q_1^{[\ell]} \prec M q_2^{[\ell]}
$$
where the notation $\bullet^{[k]}$ denotes the first $k$ rows of a matrix/vector. 
\end{enumerate}
\end{lemma}
Note that $T_i^{h}$ is of full row rank for $i=1,2$ since $h=\min\{h_1,h_2\}$. Then, it follows from \eqref{e:T-h} and Lemma~\ref{le:04} that either
\beq\label{e:T12-h}
T_1^{h}=MT_2^{h}\quad\text{and}\quad {\bf e}_1^h=M{\bf e}_2^h
\eeq
for some $(h+1)\times(h+1)$ lower triangular matrix $M$ with positive diagonal elements, or there exist $\ell$ with $1\leqslant \ell \leqslant h+1$ and $\ell\times \ell$ lower triangular matrix $M$ with positive diagonal elements such that  
\beq
T_1^{\ell-1} = M T_2^{\ell-1}\quad\text{and}\quad{\bf e}_1^{\ell-1} \succ M {\bf e}_2^{\ell-1}.
\eeq
Suppose first that the latter holds. Since $T_1^0=c^T=T_2^0$ and ${\bf e}_1^{0}=f={\bf e}_2^{0}$, the case $\ell=1$ is impossible. Then, we get
$$
T_1^{k} = M T_2^{k}\quad\text{and}\quad{\bf e}_1^{k} \succ M {\bf e}_2^{k}
$$
for some $k$ with $1\leqslant k \leqslant h$ and $(k+1)\times (k+1)$ lower triangular matrix $M$ with positive diagonal elements. This is nothing but the statement \ref{t:main1:s4}. Thus, it remains to show that \eqref{e:T12-h} also implies the statement \ref{t:main1:s4} or \ref{t:main1:s4.5} or \ref{t:main1:s5}. To do so, we consider two cases: $h_1 =h_2$ and $h_1\ne h_2$.

For the case $h_1\ne h_2$, we prove that \eqref{e:T12-h} implies statement \ref{t:main1:s4.5}.
Note first that the first part of statement  \ref{t:main1:s4.5}  holds due to \eqref{e:T12-h}. For the second part, we only need to prove for $h_1 <h_2$. For the case $h_1 >h_2$, the proof is similar. 

Since $h_1 <h_2$ and $T_2^{h_2}$ is of full row rank, there exists $\xi$ such that 
$T_2^{h_1}\xi +{\bf e}_2^{h_1}=0 \text{ and } T_2^{h_2}\xi +{\bf e}_2^{h_2}\succ 0.$
Together with \eqref{e:T-h+1.2} and \eqref{e:T12-h}, this implies
$$
T_1^{h_1}\xi + {\bf e}_1^{h_1}=0 \text{ and }  T_1^{h_1+1}\xi +{\bf e}_1^{h_1+1} \succ 0.
$$
This immediately implies $c^TA_1^{h_1}e_1 - p_1^T{\bf e}_1^{h_1}>0.$

For the case $h_1 =h_2=h$, we prove that \eqref{e:T12-h} implies the statement \ref{t:main1:s4} or \ref{t:main1:s5}. To this end, first note that it suffices to show that either
\beq
T_1^{h+1}=\hatM T_2^{h+1}\quad\text{and}\quad {\bf e}_1^{h+1}\succ\hatM{\bf e}_2^{h+1}
\eeq
or
\beq
T_1^{h+1}=\hatM T_2^{h+1}\quad\text{and}\quad {\bf e}_1^{h+1}=\hatM{\bf e}_2^{h+1}
\eeq
for some $(h+2)\times(h+2)$ lower triangular matrix $\hatM$ with positive diagonal elements. Indeed, the former would imply the statement \ref{t:main1:s4} and the latter \ref{t:main1:s5}. In what follows, we will construct a matrix $\hatM$ which will satisfy one of these two conditions. To do so, note that 
\beq\label{e:ca-h}
c^TA_i^{h+1} = p_i^T T_i^{h}\text{ with }i=1,2
\eeq
for some $p_1, p_2\in \R^{h+1}$ as $h$ is the observability index of both $(c^T,A_1)$ and $(c^T,A_2)$. Define
$$
q^T=p_1^TM-\alpha p_2^T\quad\text{and}\quad\hatM=\bbm M & 0\\q^T & \alpha\ebm
$$
for some $\alpha$. It follows from \eqref{e:T12-h} and \eqref{e:ca-h} that $T_1^{h+1}=\hatM T_2^{h+1}$. Therefore, it remains to show that we can choose $\alpha>0$ such that either ${\bf e}_1^{h+1}\succ\hatM{\bf e}_2^{h+1}$ or ${\bf e}_1^{h+1}=\hatM{\bf e}_2^{h+1}$. Since ${\bf e}_1^{h}=M{\bf e}_2^{h}$, ${\bf e}_1^{h+1}\succ\hatM{\bf e}_2^{h+1}$ holds if and only if $c^TA_1^he_1>q^T{\bf e}_2^h+\alpha c^TA_2^he_2$ and ${\bf e}_1^{h+1}=\hatM{\bf e}_2^{h+1}$ holds if and only if $c^TA_1^he_1=q^T{\bf e}_2^h+\alpha c^TA_2^he_2$. As such, it is enough to show that we can choose $\alpha>0$ such that $c^TA_1^he_1\geq q^Te_2^h+\alpha c^TA_2^he_2$. By using the definition of $q$, we see that the last inequality is equivalent to 
\beq\label{e:inq-alpha}
(c^TA_1^he_1-p_1^T{\bf e}_1^h)\geq \alpha (c^TA_2^he_2-p_2^T{\bf e}_2^h).
\eeq
Since $T_1^h$ is of full column rank, there exists $\xi_0$ such that $T_1^h\xi_0+{\bf e}_1^h=0$. It follows from \eqref{e:T12-h} that $T_2^h\xi_0+{\bf e}_2^h=0$. By using \eqref{e:ca-h}, we can rewrite \eqref{e:inq-alpha} as
\beq
(c^TA_1^he_1+c^TA_1^{h+1}\xi_0)\geq \alpha (c^TA_2^he_2+c^TA_2^{h+1}\xi_0).
\eeq
Define $\rho_i=c^TA_i^he_1+c^TA_i^{h+1}\xi_0$ for $i=1,2$ and observe that there exists $\alpha>0$ satisfying the above inequality unless $(\rho_1\leq 0\text{ and }\rho_2> 0)$ or $(\rho_1<0\text{ and }\rho_2\geq 0)$. However, neither of these two cases can occur due to the definition of $\xi_0$ and \eqref{e:T-h+1} with $\xi=\xi_0$.

\subsubsection{Proof of Theorem \ref{t:main1}.\ref{t:main1-b}}
On the one hand, right-uniqueness of Filippov solutions (statement \ref{t:main1:s1}) necessitates that the implication
\beq\label{e:on-w0}
\dot x_i = A_i x_i + e_i,\, x_i(0) = \xi\implies x_1(t) = x_2(t)\,\, \text{ for all }  t\geqslant 0
\eeq
holds for all $\xi\in \mathcal W_1^0 \cap \mathcal W_2^0$. On the other hand, it follows from statement \ref{t:main1:s5} and \eqref{e:prop:wch1} that $\calW_1^0=\calW_2^0=\set{\xi}{T_1^{h+1}\xi+{\bf e}_1^{h+1}=0}$. As such, statement \ref{t:main1:s6} immediately follows from \eqref{e:on-w0}.

\subsubsection{Proof of Theorem \ref{t:main1}.\ref{t:main1-c}}\label{s:imp-c}
Note that both statements \ref{t:main1:s5} and \ref{t:main1:s6} are invariant under time-reversal. As such, it is enough to show that the forward Carath\'eodory property holds for every Filippov solution.

It follows from statement \ref{t:main1:s5} that $\calW^0:=\calW_1^0=\calW_2^0$. We claim first that forward Carath\'eodory property holds for every Filippov solution with the initial state $\xi\in\calW^0$. To this end, note that statement \ref{t:main1:s6} implies that
\beq\label{e:w0-imp-inv}
\xi\in\calW^0\implies A_1\xi+e_1=A_2\xi+e_2.
\eeq
Note also that the implication
\beq\label{e:wi0-imp}
\xi\in\calW_i^0\implies A_i\xi+e_i\in\calW_i^0
\eeq
readily holds from the very definition of the sets $\calW_i^0$ for $i=1,2$. Together with \eqref{e:w0-imp-inv}, this invariance property yields
\beq\label{e:on-hyper}
\dot x_i = A_i x_i + e_i,\, x_i(0) = \xi\implies x_1(t) = x_2(t)\,\, \text{ for all } t\in\R
\eeq
for all $\xi\in \mathcal W^0$. Let $x^*$ satisfy $\dot x^* = A_1 x^* + e_1$ with $x^*(0)=\xi$. Also let $x$ be a Filippov solution of the system with $x(0)=\xi$. Then, there exists a function $\lambda:\R\rightarrow[0,1]$ such that
\beq
\dot x(t)=\lambda(t)[A_1x(t)+e_1]+\big(1-\lambda(t)\big)[A_2x(t)+e_2]
\eeq
for almost all $t\in\R$. It follows from \eqref{e:on-hyper} that
\beq
\dot x^*(t)=\lambda(t)[A_1x^*(t)+e_1]+\big(1-\lambda(t)\big)[A_2x^*(t)+e_2]
\eeq
for all $t\in\R$. Define $A(t):=\lambda(t) A_1 +(1-\lambda(t))A_2$. Then, we  get
\begin{align*}
\dfrac{d}{dt}(||x^*(t)-x(t)||^2) &= 2 \big \langle x^*(t)-x(t), \dot x^*(t) - \dot x(t) \big\rangle \\&= 2 \big \langle x^*(t)-x(t),A(t)( x^*(t) -  x(t)) \big\rangle   \leqslant \alpha ||x^*(t)-x(t)||^2 
\end{align*}
where $\alpha :=2\max\{\norm{\lambda A_1 +(1-\lambda)A_2}\ | \ \lambda \in [0,1]\}.$ Since $x^*(0)-x(0)=0$, the last inequality readily implies that $x(t)=x^*(t)$ for all $t\in\R$. In other words, $x^*$ is the unique Filippov solution for the initial state $\xi$. It follows from \eqref{e:wi0-imp} that $x^*(t)\in\calW^0$ for all $t\in\R$. Hence, $x^*$ is a forward Carath\'eodory solution and statement \ref{t:main1:s3} holds for every Filippov solution with the initial state $\xi\in\calW^0$.

Next, we claim that if $x$ is a Filippov solution for the initial state $\xi$ and $t^*\in\R$ is such that $x(t^*)\in\calW^0$ then $x(t)\in\calW^0$ for all $t\in\R$. To see this, note that $\barx(t)=x(t+t^*)$ is a Filippov solution for the initial state $x(t^*)\in\calW^0$. As such, the above argument yields $x(t)=\barx(t-t^*)\in\calW^0$ for all $t\in\R$.

Therefore, it remains to show that forward Carath\'eodory property holds for every Filippov solution $x$ with the property that 
\beq\label{e:not-in-w0}
x(t)\not\in\calW^0 
\eeq
for all $t\in\R$. Let $x$ be such a Filippov solution and let $t^*\in\R$. If $c^Tx(t^*)+f\neq 0$, it follows from statement \ref{t:main1:s5} and continuity of $x$ that there exists $\epsilon_{t^*}>0$ such that \eqref{eq:fc} holds. Suppose that $c^Tx(t^*)+f=0$. It follows from \eqref{e:not-in-w0} and \eqref{e:prop:wch1} that there exists an integer $k$ with $0\leq q\leq h$ such that
\begin{align}
c^TA_1^{\ell}x(t^*)+c^TA_1^{\ell-1}e_1&=0\text{ for all }\ell=1,2,\ldots,q\label{e:a1-q}\\
c^TA_1^{q+1}x(t^*)+c^TA_1^{q}e_1&\neq0.
\end{align}
Now, suppose that 
\beq\label{e:a1-q-2}
c^TA_1^{q+1}x(t^*)+c^TA_1^{q}e_1>0.
\eeq
It follows from statement \ref{t:main1:s5} that
\begin{align}
c^TA_2^{\ell}x(t^*)+c^TA_2^{\ell-1}e_2&=0\text{ for all }\ell=1,2,\ldots,q\label{e:a2-q}\\
c^TA_2^{q+1}x(t^*)+c^TA_2^{q}e_2&>0\label{e:a2-q-2}.
\end{align}
Since $x$ is continuous, there must exist $\epsilon_{t^*}>0$ such that
\begin{align}
c^TA_1^{q+1}x(t)+c^TA_1^{q}e_1&>0\\
c^TA_2^{q+1}x(t)+c^TA_2^{q}e_2&>0
\end{align}
for all $t\in (t^*,t^*+\epsilon_{t^*})$.

For $\lambda \in [0, 1]$, we define $ A(\lambda) := \lambda A_1 + (1-\lambda)A_2, \ e(\lambda):= \lambda e_1 +(1-\lambda)e_2$. We also define
\begin{gather*}
 \mathbb G_0 = \mathbb H_0 := \{I\},\\
\mathbb G_k :=\big \{A_i G'\ | \   G' \in \mathbb G_{k-1}, i=1,2  \big \}\text{ for }k\geq 1,\\
\mathbb H_k :=\big \{A(\lambda) H' \ |  \  \lambda \in [0, 1]\  \text {and} \ H' \in \mathbb H_{k-1}  \big \} \text{ for }k\geq 1.
\end{gather*}
For each $k\geqslant 1$ and  $(G,H)\in \mathbb G_k\times\mathbb H_k$ of the form
\begin{gather*}
G= A_{i_k}A_{i_{k-1}}\cdots A_{i_{1}},\  H = A(\lambda_k)A(\lambda_{k-1})  \cdots A(\lambda_1),
\end{gather*}
we define  
\begin{gather*}
e_G := c^T A_{i_k}A_{i_{k-1}}\cdots A_{i_{2}}e_{i_1} \text{ and } e_H := c^T A(\lambda_k) A(\lambda_{k-1}) \cdots A(\lambda_2)e(\lambda_1)
\end{gather*}
with the convention that $e_I:= f$ for $k=0$. Finally, define
 $ \mathcal G_k :=\{ (G, e_G) \ | \ G\in \mathbb G_k\}$ and $ \mathcal H_k :=\{ (H, e_H) \ | \ H\in \mathbb H_k\}.$

Note that $\mathbb G_k \subseteq \mathbb H_k$ and $\mathcal G_k \subseteq \mathcal H_k $ for all $k \geqslant 0$. Also note that $\conv(\mathcal G_k)\subseteq \conv (\mathcal H_k)$ and that $\mathcal H_k\subseteq \conv(\mathcal G_k)$. Hence, we have 
\beq\label{e:lm415}
\conv (\mathcal H_k) = \conv (\mathcal G_k)
\eeq
for every $k\geqslant 0$.

Now, we claim that for each $\ell\in\pset{0,1,\ldots,q}$ and each $G\in\mathbb G_\ell$
\beq\label{e:g-zero}
c^TGx(t^*)+e_G=0.
\eeq
To prove this claim, we make an induction on $\ell$. The case $\ell=0$ is evident. Suppose that the claim holds for all $\ell$ with $0\leq\ell<p\leq q$. Let $G\in\mathbb G_p$. In the cases $G=A_1^p$ and $G=A_2^p$, the claim readily follows from \eqref{e:a1-q} and \eqref{e:a2-q}, respectively. Otherwise, $G=A_{i}^{k_1}A_{3-i}^{k_2}G'$ for some $i\in \{1,2\}$, $2\leq k_1+k_2\leq p$, and $G'\in\mathbb G_{p-k_1-k_2}$. Then, it follows from statement \ref{t:main1:s5} that there exists a positive number $\alpha$ such that
\beq
c^TGx(t^*)+e_G=\alpha(c^T\barG x(t^*)+e_{\barG})
\eeq
where $\barG=A_{i}^{k_1+k_2}G'$. By repeating the same argument, one can prove the existence of a positive number $\alpha_G$ such that
\beq
c^TGx(t^*)+e_G=\alpha(c^TA_i^p x(t^*)+c^TA_i^{p-1}e_i).
\eeq
Then, the claim follows from either \eqref{e:a1-q} or \eqref{e:a2-q}. The very same argument employed in the last step of the above induction yields
\beq
c^TGx(t^*)+e_G>0
\eeq
for all $G\in \mathbb G_{q+1}$. Since $x$ is continuous and $\mathbb G_{q+1}$ is a finite set, we can conclude that there exists a positive number $\epsilon^*$ such that
\beq\label{e:g-greater}
c^TGx(t)+e_G>0
\eeq
for all $G\in \mathbb G_{q+1}$ and for all $t\in (t^*,t^*+\epsilon^*)$.

From \eqref{e:lm415}, we further get
\beq\label{e:htstar}
c^THx(t^*)+e_H=0
\eeq
for all $H\in\mathbb H_\ell$ and $\ell=0,1,\ldots,q$, and also that
\beq\label{e:ht-greater}
c^THx(t)+e_H>0
\eeq 
for all $H\in\mathbb H_{q+1}$ and for all $t\in (t^*,t^*+\epsilon^*)$.

Let $H\in\mathbb H_{q}$. Note that
\beq\label{e:hq+1}
c^TH\dot{x}(t)\in\set{c^TH'x(t)+e_{H'}}{H'\in\mathbb H_{q+1}}.
\eeq
Also note that
\beq
c^THx(t)+e_{H}=c^THx(t^*)+e_{H}+\int_{t^*}^tc^TH\dot{x}(s)\,ds.
\eeq
Then, it follows from \eqref{e:htstar}, \eqref{e:ht-greater}, and \eqref{e:hq+1} that 
\beq
c^THx(t)+e_{H}>0
\eeq
for all $t\in (t^*,t^*+\epsilon^*)$. The very same argument can be repeated for $H\in\mathbb H_{\ell}$ with $\ell=q-1,\ldots,1,0$ to obtain
\beq
c^THx(t)+e_{H}>0
\eeq
for all $H\in\mathbb H_\ell$, $\ell=0,1,\ldots,q$, and $t\in (t^*,t^*+\epsilon^*)$. In particular, we obtain
\beq
c^Tx(t)+f>0
\eeq
for all $t\in (t^*,t^*+\epsilon^*)$ with the choice of $\ell=0$. Therefore, $x$ is a forward Carath\'eodory solution.

The case
\beq
c^TA_1^{q+1}x(t^*)+c^TA_1^{q}e_1<0
\eeq
can be proven by using the above arguments in a similar fashion.

\subsubsection{Proof of Theorem \ref{t:main1}.\ref{t:main1-d}}

As we have just shown, statements \ref{t:main1:s5} and \ref{t:main1:s6} imply that every Filippov solution is a forward Carath\'eodory solution. Then, it is enough to show that every forward Carath\'eodory solution is right-unique. 

Let $x_1, x_2$ be two forward Carath\'eodory solutions with $x_1(0) = x_2(0) =\xi$. Then, there exists  $\epsilon >0$ such that 
\bse\label{e:fwc}
\begin{align}
\dot x_1 (t)&=  A_i x_1(t) +e_i, \ (-1)^i[ c^T x_1(t) +f] \geqslant 0 \label{fwc:01}
\\
\dot x_2 (t)&=  A_j x_2(t) +e_j, \ (-1)^j[ c^T x_2(t) +f] \geqslant 0 \label{fwc:02}
\end{align}
\ese
for all $t\in [0, \epsilon)$. In case $i=j$, we have readily $x_1(t) = x_2(t)$ for all $t\in [0,\epsilon)$. In case $i\ne j$, we can assume, without loss generality, that $i=1$ and $j=2$. Then, it follows from \eqref{e:fwc} that $\xi \in \mathcal W_1$ and $\xi \in \mathcal W_2$. In other words, we have
\beq
(-1)^i\bigg (\begin{bmatrix}c^T\\ c^TA_i \\ \vdots \\  c^TA_i^{h+1}\end{bmatrix}\xi+\begin{bmatrix} f\\ c^Te_i \\ \vdots \\ c^TA_i^{h}e_i\end{bmatrix}\bigg)\succeq  0
\eeq
for $i\in\pset{1,2}$. Then, it follows from statement \ref{t:main1:s5} that 
\beq
\begin{bmatrix}c^T\\ c^TA_i \\ \vdots \\  c^TA_i^{h+1}\end{bmatrix}\xi+\begin{bmatrix} f\\ c^Te_i \\ \vdots \\ c^TA_i^{h}e_i\end{bmatrix}= 0
\eeq
for $i\in\pset{1,2}$ and hence $\xi\in\calW^0=\calW_1^0=\calW_2^0$. As proven in \ref{s:imp-c}, this means that $x_1(t)=x_2(t)=x^*(t)$ for all $t\in\R$ where $x^*$ satisfies $\dot{x}^*=A_1x^*+e_1$ with $x^*(0)=\xi$. 

\subsection{Proof of Theorem~\ref{thrm:main3}}

Since $M$ is a lower triangular matrix with positive diagonal elements, it can be partitioned as
$$M =\begin{bmatrix} M_1&0\\ *&m_{33} \end{bmatrix} = \begin{bmatrix} m_{11}&0 &0 \\ *&m_{22}&0 \\ *& * & m_{33} \end{bmatrix}$$
where $m_{ii}>0$ for $i=1,\ldots,3$. 

It follows from \eqref{eq:thrm:main3-0} that either
\begin{equation}
\begin{bmatrix}c^T\\ c^TA_1\end{bmatrix} = M_1 \begin{bmatrix}c^T\\ c^TA_2\end{bmatrix}, \begin{bmatrix}f\\ c^Te_1\end{bmatrix} \succ M_1 \begin{bmatrix}f\\ c^Te_2\end{bmatrix}
\end{equation}
or
\begin{equation}\label{eq:pthrm:main3-1}
\begin{bmatrix}c^T\\ c^TA_1\end{bmatrix} = M_1 \begin{bmatrix}c^T\\ c^TA_2\end{bmatrix}, \begin{bmatrix}f\\ c^Te_1\end{bmatrix} =M_1 \begin{bmatrix}f\\ c^Te_2\end{bmatrix}
\end{equation}
holds. The claim follows readily for the former case due to Theorem~\ref{thrm:main2}. Therefore, it remains to prove the claim when \eqref{eq:pthrm:main3-1} holds.

Let $\xi\in \R^n\backslash \Omega$. Due to \eqref{eq:thrm:main3-0}, \eqref{eq:pthrm:main3-1} and the definition of $\Omega$, only the following cases are possible: 
\begin{enumerate}
\item $c^T\xi +f \ne 0$.
\item $c^T\xi +f =0, (c^TA_1\xi +c^Te_1)(c^TA_2\xi +c^Te_2)>0.$
\item $c^T\xi +f=c^TA_1\xi + c^Te_1 =c^TA_2\xi + c^Te_2=0 \text{ and }$
$$(c^TA_1^2\xi +c^TA_1e_1)(c^TA_2^2\xi +c^TA_2e_2)>0. $$
\item $c^T\xi +f =c^TA_1\xi +c^Te_1=c^TA_2\xi +c^Te_2=0 \text { and }$
\begin{equation}\label{eq:lm4.15-2}
c^TA_1^2\xi+c^TA_1e_1 >0, c^TA_2^2\xi+c^TA_2e_2 <0.
\end{equation}
\end{enumerate}
By similar arguments to those in the proof of Theorem~\ref{t:main1}.\ref{t:main1-c}, one can conclude the right-uniqueness of Filippov solutions for the initial state $\xi$ for the first three above-mentioned cases.

For the last case, we claim that if $x$ is a Filippov solution with $x(0)=\xi$ then there exists $\epsilon >0$ such that
\begin{subequations} \label{eq:cl}
\begin{equation}
c^T x(t) + f=0
\end{equation}
\begin{equation}
 c^TA_1x(t)+c^Te_1 =c^TA_2x(t)+c^Te_2 =0
\end{equation}
\end{subequations}
for all $t\in [0, \epsilon)$. Note that the conditions \eqref{eq:cl} would imply that $x$ satisfies the differential inclusion
\begin{align}\label{eq:inc-03}
\dot x \in \begin{cases}\{A_1x + e_1\} &\text{if} \ \tilde c^Tx + \tilde f <0\\ \conv\{A_1x + e_1, A_2x+e_2\} &\text{if} \ \tilde c^Tx + \tilde f =0\\  \{A_2x + e_2\} &\text{if} \ \tilde c^Tx + \tilde f >0\end{cases}
\end{align}
on the interval $[0,\epsilon)$ where $\tilde c^T =c^T A_1$ and $\tilde f = c^T e_1$. As such, right-uniqueness of $x$ would follow from \eqref{eq:lm4.15-2} and \cite[Theorem 2.10.2]{filippov:88} since $\tilde c^T \xi + \tilde f =0 $ and $\tilde c^T A_1 \xi + \tilde c^T e_1 >0$.

Therefore, it is enough to prove the existence of $\epsilon>0$ such that the conditions \eqref{eq:cl} are satisfied. To do so, we first note that one can find positive numbers $\delta$ and $\epsilon$ such that
\begin{equation}\label{eq:lm4.15-3}
c^TA_1^2x(t) +c^TA_1e_1 \geqslant \delta, \quad c^TA_2^2x(t) +c^TA_2e_2 \leqslant -\delta 
\end{equation}
for all $t\in [0,\epsilon)$ due to \eqref{eq:lm4.15-2}, $x(0) =\xi$ and the continuity of $x$.
To show that the conditions \eqref{eq:cl} are satisfied with this choice of $\epsilon>0$, consider the functions $V_1, V_2:\R^n\to \R$ defined by
$$
V_1(z) =\begin{cases} (c^T z+f)(c^TA_1^2z+c^TA_1e_1) & \text{if} \ c^T z+f \leqslant 0\\ (c^T z+f)(c^TA_2^2z+c^TA_2e_2) & \text{if} \ c^Tz +f \geqslant 0, \end{cases}
$$
$$
V_2(z) =\begin{cases} c^TA_1z+c^Te_1 & \text{if} \  c^T z+f \leqslant 0\\ m_{22}(c^TA_2z+c^Te_2) & \text{if} \ c^T z+f \geqslant 0.\end{cases}
$$
Also consider
$V(t):= \dfrac{1}{2}V^2_2(x(t))-V_1(x(t)).$ From \eqref{eq:lm4.15-3}, we get 
\begin{equation} \label{eq:lm4.15-5}
V(t) \geqslant \delta |c^T x(t) +f| + \dfrac{1}{2}V_2^2(x(t))\geqslant 0 
\end{equation}
for all $t\in [0,\epsilon)$. Note that the relation
\begin{equation}\label{eq:lm4.15-6}
\dot V(t) \leqslant \lambda V(t)
\end{equation} 
for almost all $t\in [0,\epsilon)$ and for some $\lambda >0$ would imply $V(t)=0$ for all $t\in [0,\epsilon)$ since $V(0)=0$. Together with \eqref{eq:lm4.15-5} and the definition of $V_2$, this would imply that the conditions \eqref{eq:cl} are satisfied. As such, it is enough to show \eqref{eq:lm4.15-6} in order to complete the proof.

Define $r:=\max\{\norm{x(t)} \ | \ t\in [0,\epsilon]\}$ and $B(0,r):=\{x \ | \ \norm{x} \leqslant r\}$. Note that $V_1$ is readily continuous and also that $V_2$ is continuous due to \eqref{eq:pthrm:main3-1}. Furthermore, both $V_1$ and $V_2$ are Lipschitz on $B(0,r)$. Together with the absolute continuity of $x$, this implies that $V_1(x(\cdot))$ and $V_2(x(\cdot))$ are even absolutely continuous on $[0,\epsilon)$.

Let $\Lambda$ be the set of all $t\in [0,\epsilon)$ for which $x$, $V_1(x(\cdot))$, and $ V_2(x(\cdot))$ are differentiable. Due to the absolute continuity of these functions, the set $[0,\epsilon)\setminus \Lambda$ is of measure zero. Thus, it suffices to prove that \eqref{eq:lm4.15-6} holds for all $t\in\Lambda$.

To do so, let $t^*\in \Lambda$. First, suppose that $c^Tx(t^*) + f \ne 0$. In this case, we get  
\begin{equation}\label{eq:lm4.15-7}
\dot V(t^*)\leqslant L |c^T x(t^*)+f| \leqslant \dfrac{L}{\delta} V(t^*)
\end{equation}
where $L:=\max\{ |c^TA_i^3x(t)+c^TA_i^2e_i|: t\in [0,\epsilon], i=1,2\}$. In other words, \eqref{eq:lm4.15-6} holds for $t^*$.

Now, suppose that $c^Tx(t^*)+f=0$. The Taylor expansion of $x$ around $t^*$ yields
\begin{equation}\label{eq:lm4.15-7b}
c^T x(t^*+\tau) +f = \tau c^T \dot x(t^*) +c^T o(\tau)
\end{equation}
for all $\tau$ sufficiently close to $0$ where $\lim_{\tau\rightarrow 0}o(\tau)/\tau=0$. It follows from \eqref{eq:lm4.15-7b} that the set of the one-sided derivatives of $V_1(x(\cdot))$ at $t^*$ is equal to 
\begin{equation} \label{eq:lm4.15-8}
\{c^T\dot x(t^*)(c^TA_i^2 x(t^*) +c^T A_i e_i) \ | \ i=1,2\}.
\end{equation}
Since $V_1(x(\cdot))$ is differentiable at $t^*$, this set must be a singleton. In view of \eqref{eq:lm4.15-3}, this can happen only if $c^T\dot x(t^*)=0$. This means that $\dot V_1(x(t^*))=0$. Since $x$ is a Filippov solution, we get  
\begin{equation}\label{eq:lm4.15-9}
0\in \conv\{c^TA_1x(t^*)+c^Te_1,c^TA_2x(t^*)+c^Te_2\}.
\end{equation}
By post-multiplying the first equation in \eqref{eq:pthrm:main3-1} by $x(t^*)$ and adding to the second, we obtain 
\begin{equation}\label{eq:lm4.15-10}
c^TA_1x(t^*)+c^Te_1 = m_{22} (c^TA_2x(t^*)+c^Te_2).
\end{equation}
From \eqref{eq:lm4.15-9} and \eqref{eq:lm4.15-10}, we get
$$ c^TA_1x(t^*)+c^Te_1 = c^TA_2 x(t^*)+c^Te_2 =0.$$
From the definition of $V_2$, this yields $V_2(x(t^*))=0$ and further
$$\dfrac{d}{dt}(V_2^2(x(t^*)))=0, \ \dot V(t^*)=\dfrac{d}{dt}(V_2^2(x(t^*)))-\dfrac{d}{dt}V_1(x(t^*))=0.$$
Clearly, \eqref{eq:lm4.15-6} holds for $t^*$.
\EP

\section{Conclusions}\label{sec:conclusions} 
We studied existence, uniqueness and nature of solutions Carath\'{e}odory and Filippov solutions for bimodal (possibly discontinuous) piecewise affine systems in a differential inclusion setting. First, we showed that the typical conditions that are employed in the context of general differential inclusions in order to guarantee uniqueness of Filippov are quite restrictive in the context of piecewise affine systems. Then, we presented a set of necessary and a set of sufficient conditions that ensure uniqueness of Filippov solutions for bimodal piecewise affine systems. By investigating the relationships between Carath\'{e}odory and Filippov under the presented condition, we provide conditions that rule our the so-called Zeno behavior. Possible extensions of the main results of this paper to general piecewise affine dynamical systems with external inputs emerge as future research directions.

\bibliographystyle{plain} 
\bibliography{referencefile}

\end{document}